\documentclass[aps,prl,twocolumn,showpacs,superscriptaddress]{revtex4-1}
\usepackage{graphicx}
\usepackage{verbatim}
\usepackage{amsmath}
\usepackage{pdfpages}

\begin{document}

\title{Measurement of the spectral line shapes for orbital excitations in the Mott insulator CoO using high-resolution resonant inelastic x-ray scattering}

\author{L. Andrew Wray}
\affiliation{Advanced Light Source, Lawrence Berkeley National Laboratory, Berkeley, CA 94720, USA}
\author{J. Li}
\author{Z. Q. Qiu}
\affiliation{Physics Department, University of California, Berkeley, CA 94720, USA}
\author{Jinsheng Wen}
\affiliation{Physics Department, University of California, Berkeley, CA 94720, USA}
\affiliation{Condensed Matter Physics and Materials Science Department, Brookhaven National Laboratory, Upton, NY 11973, USA}
\author{Zhijun Xu}
\author{Genda Gu}
\affiliation{Condensed Matter Physics and Materials Science Department, Brookhaven National Laboratory, Upton, NY 11973, USA}
\author{Shih-Wen Huang}
\author{Elke Arenholz}
\author{Wanli Yang}
\author{Zahid Hussain}
\author{Yi-De Chuang}
\affiliation{Advanced Light Source, Lawrence Berkeley National Laboratory, Berkeley, CA 94720, USA}

\pacs{78.47.je, 71.27.+a, 78.70.Ck, 31.15.-p}

\begin{abstract}

We establish the spectral line shape of orbital excitations created by resonant inelastic X-ray scattering for the model Mott insulator CoO. Improved experimental energy resolution reveals that the line shapes are strikingly different from expectations in a first principles-based atomic multiplet model. Extended theoretical simulations are performed to identify the underlying physical origins, which include a pronounced thermal tail reminiscent of anti-Stokes scattering on the energy gain side of excitations, and an essential contribution from interatomic many-body dynamics on the energy loss side of excitations.

\end{abstract}


\maketitle

\section{Introduction}
Many types of many-body phases and dynamics derive from low energy spin, lattice, phase and orbital degrees of freedom that are challenging to excite and resolve in experiments. Resonant inelastic X-ray scattering (RIXS) is a powerful experimental technique for observing many-body collective excitations \cite{Kuiper,spinFlipTheory,SpinFlipObsLedge,GhirEarlyRIXS,Hasan_KRIXS,NiO_MRIXS,VDBultrashort,PattheyML,VDBmagneticLCO,AmentRIXSReview,SalaLedge,DevereauxNew,spinOrbitalSep,deGrootNewCoO,WrayNiO}, but has been limited by energy resolution $\delta$E$\gtrsim$0.1eV that is much larger than room temperature thermal fluctuations ($k_B$$\times$300K=26meV). In the absence of better resolution, first principles-based atomic multiplet simulations that neglect most many-body dynamics have achieved reasonably good agreement with experimental direct RIXS spectra \cite{spinFlipTheory,SpinFlipObsLedge,NiO_MRIXS,PattheyML,AmentRIXSReview,SalaLedge,deGrootNewCoO,WrayNiO}. For this study, improved energy resolution of $\delta$E$\sim$30meV is obtained by using low energy photons at the cobalt M-edge (50eV$\leq$h$\nu$$\leq$70eV). With this resolution, clear discrepancies between atomic multiplet excitations and experimental features in model Mott insulator CoO are revealed. Features on the energy gain side of excitations are shown to match a thermal effect present in multiplet models, referred to here as psuedo-anti-Stokes (pAS) scattering. Features on the energy loss side are attributed to intersite interactions between multiplet excitations and low energy many-body degrees of freedom.

\begin{figure}[t]
\includegraphics[width = 8.7cm]{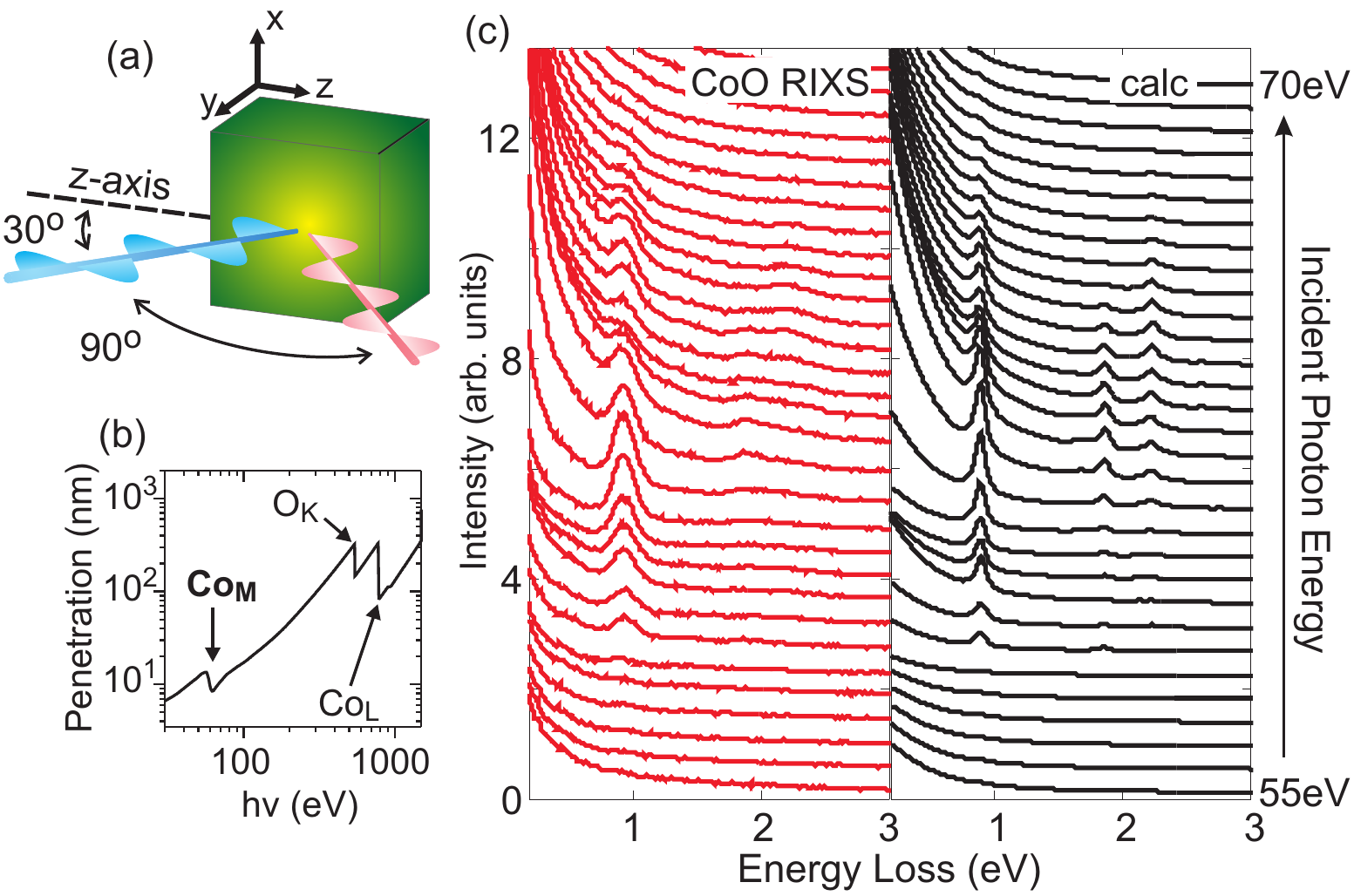}
\caption{{\bf{Resonant inelastic scattering at the M-edge}}: (a) The scattering geometry is shown. (b) The penetration length of X-rays in CoO is estimated from Ref. \cite{XrayPath}. (c) (left) CoO RIXS spectra measured at 320K with incident energies from 55-70eV (0.5eV step) are compared with (right) an atomic multiplet model.}
\end{figure}

\section{Materials and Experimental Methods}

CoO is a large spin antiferromagnet with a cubic NaCl crystal structure \cite{cubicNote,SM} and predominantly 3d$^7$ cobalt valence \cite{ElkeXLD,deGrootNewCoO}. The type 2 Neel transition occurs at roughly 290K \cite{CoOtype2}, however magnetic excitation peaks observed by inelastic neutron scattering do not soften significantly over a temperature range extending to above 320K \cite{CoOinelastNeutron}, and short range antiferromagnetic correlation persists at a length scale of $>$$100\AA$ at 310K \cite{CoOHTcorr}.

For this study, a bulk-like well ordered CoO film with single-nanometer surface roughness was grown on MgO, and transferred to the RIXS chamber vacuum with minimal exposure to the air. The sample was grown by molecular beam epitaxy (MBE) in an ultrahigh vacuum chamber with the base pressure of 2$\times$10$^{-10}$ torr. Cobalt was evaporated onto a MgO(001) substrate in 2$\times$10$^{-6}$ torr oxygen at room temperature. After evaporation, the sample was annealed to 600$^\circ$C within a 2$\times$10$^{-6}$ torr oxygen atmosphere for 30 minutes, yielding a surface roughness under 1nm measured by atomic force microscopy. Sample thickness was 45nm, much greater than the $\lesssim$5nm layer over which interface strain and exchange effects may be significant \cite{thicknessNeel}, and similar to three times the estimated X-ray penetration depth (Fig. 1(b)). M-edge measurements were performed with $\sim$30meV energy resolution at the beamline 4.0.3 (MERLIN) RIXS endstation (MERIXS) \cite{YiDeDetector} at the Advanced Light Source (ALS), Lawrence Berkeley National Laboratory. The scattering geometry is shown in Fig. 1(a). The ratio of resonant to non-resonant scattering rate ($A_R/A_{NR}$) can be determined from the x-ray penetration depth plot in Fig. 1(b), where the depth of each resonant dip is proportional to $\log(1+A_R/A_{NR})$. The shallow M-edge dip suggests that resonant self-absorption will have less influence on spectra at the M-edge relative to higher energy resonances.

\section{Temperature dependence in multiplet RIXS excitations}

The incident energy dependence of RIXS spectra for CoO is presented in Fig. 1(c, left), together with an atomic multiplet simulation (right panel). The prominent feature at 0.8-1.1eV energy loss is associated with the $^4T_2$ symmetry multiplet states, which are distinguished from lower energy states by a larger charge density in e$_g$ orbitals. All modeled energy loss features are convoluted by a 70meV Lorentzian function to represent intrinsic lifetime broadening.

The simulation follows the standard Kramers-Heisenberg method that accurately describes CoO RIXS at the L-edge \cite{deGrootNewCoO} and NiO at the M-edge \cite{NiO_MRIXS,WrayNiO}, with a renormalized first principles parameter set \cite{detailedVals, SlaterCondonParams}. The crystal field perturbation of 10Dq=0.98eV (at T=90K) is fixed by the RIXS excitation energies and Tanabe-Sugano diagram for 3d$^7$ cobalt, and reduced by 10$\%$ in the intermediate states due to orbital contraction. The unit cell volume contracts from V$_{320K}$=77.4[1] \AA$^3$ at 320K to V$_{90K}$=76.5[5] \AA$^3$ at 90K \cite{CoOlatParam}, giving a change of 1-(V$_{320K}$/V$_{90K}$)$^{5/3}$=-1.9$\%$ in the estimated value of 10Dq \cite{expansion10Dqform}. An external exchange field of $J^*=0.0126eV$ is taken from neutron scattering measurements \cite{CoOinelastNeutron,CoOHTcorr}, and calculated spectra are summed over all 3 [100] magnetic domain configurations.

\begin{figure}[t]
\includegraphics[width = 8.7cm]{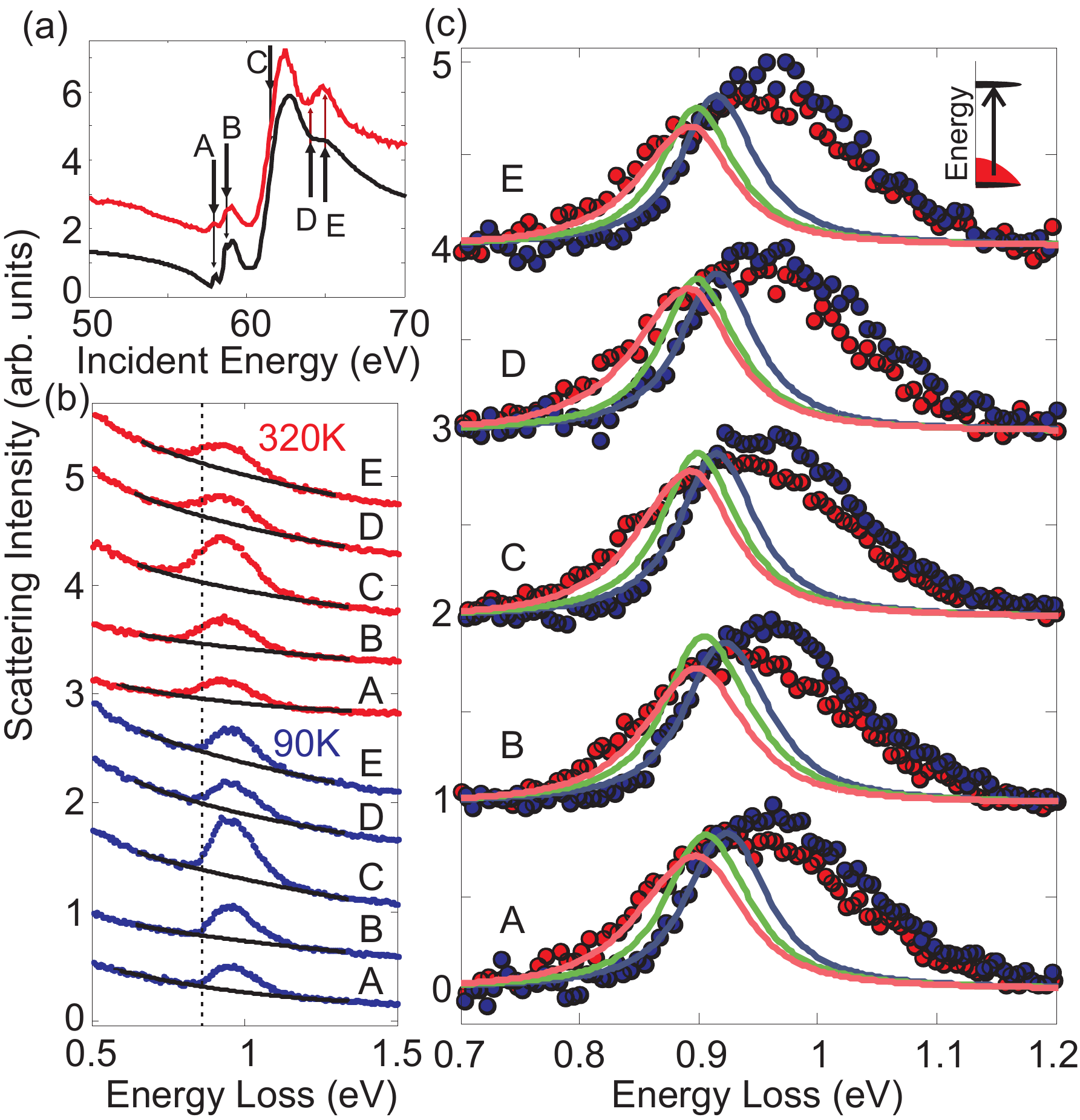}
\caption{{\bf{Thermal changes in experimental line shapes}}: (a) (red) A fluorescence spectrum of CoO is compared with (black) an atomic multiplet simulation. (b) RIXS data obtained at energies labeled A-E in panel (a) are overlaid with black background curves obtained by an automated fitting function described in the text. Red data points are measured at 320K and blue points at 90K, and a dashed vertical line indicates the onset of $^4T_2$ excitation intensity at 90K. (c) The data from panel (b) are shown after background subtraction, and normalized so that all 90K measurements have the same height. Curves represent multiplet simulations for (dark blue) T=90K, (light red) T=320K, and (green) T=90K electronic Boltzmann occupation with the T=320K lattice constant, and are normalized to align the blue simulation curve with the low energy lineshape of T=90K data points. An inset at upper right illustrates how thermally populated states (shaded red) reduce the excitation energy gap.}
\end{figure}

Core hole lifetimes are calculated using the SCLC method \cite{WrayNiO,KotaniIdea}, with a 3d-3p transition rate C$_V$ reduced by 10$\%$ relative to that estimated for the M-edge of NiO \cite{WrayNiO}, and a slow rate factor for non-multiplet decays C$_{CC}$=0.15eV. These lifetime estimates accurately reproduce detailed incident energy dependence of RIXS spectra such as the sudden appearance of an inelastic peak from 57.5-58eV in Fig. 1(c), and the narrow fluorescence peaks around 57-60eV in Fig. 2(a). The method described in Ref. \cite{WrayNiO} has been used to simulate coherent resonant elastic emission seen in Fig. 2(a) and the strong elastic tails in Fig. 1(c, right).

A pronounced softening occurs on the leading edge of RIXS intensity when temperature is raised from 90K to 320K (Fig. 2(b)). To observe the $^4T_2$ feature more clearly, background curves for the elastic tail (Fig. 2(b,black)) are obtained with an automated fitting function. The function approximates the background by fitting the local tangents and radius of curvature of a circular contour from data points at energies above and below the excitation. Examining the background subtracted curves in Fig. 2(c) reveals that the prominent $\sim$1eV $^4T_2$ feature shifts to lower energy at high temperature, the slope of the energy gain intensity onset becomes much shallower, and RIXS peak heights drop by 10-20$\%$. At both high and low temperature, the energy gain side of the peak has a sharper slope than the energy loss side. At 90K, the leading edge closely resembles a 70meV Lorentzian, while full peak width at half maximum is $\sim$160meV. The irregular shape near 1eV energy loss is suggestive of multiple closely spaced features with a continuum-like tail towards higher energy.

\begin{figure}[t]
\includegraphics[width = 8.7cm]{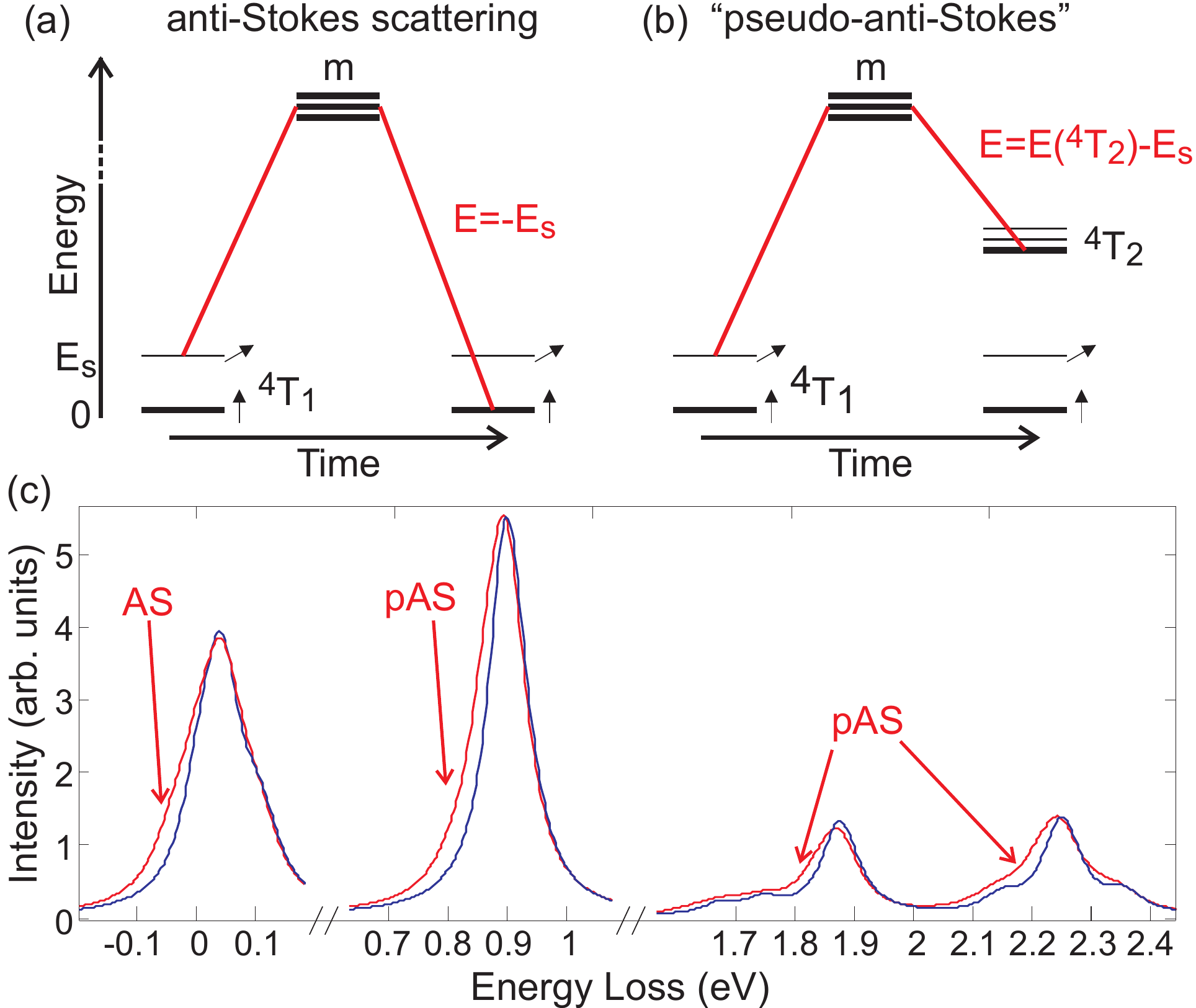}
\caption{{\bf{Pseudo-anti-Stokes features from thermally populated modes}}: (a) A schematic shows a typical anti-Stokes scattering process. Thermal fluctuations cause the scattering site to start in a spin flip excited state \cite{spinFlipTheory,NiO_MRIXS,SpinFlipObsLedge} with energy $E_S$, within the $^4T_1$ ground state manifold. An incident photon excites the system into intermediate resonance states (m), and a scattered photon is emitted as the system decays into the lowest energy state of the ground state manifold. The resulting RIXS feature appears at negative energy loss ($E=-E_S$). (b) In a pseudo-anti-Stokes scattering process, the system evolves through the same initial and intermediate states as anti-Stokes scattering, but the final state is the lowest energy state of a higher energy multiplet manifold (e.g. $^4T_2$). (c) Spectral line shapes of low energy multiplet excitations are compared at (blue) T=90K and (red) T=320K using the same 10Dq value. Anti-Stokes and pseudo-anti-Stokes tails appear on the energy gain side of all multiplet features. The spectra are averaged over incident energy, and the T=90K curve intensity has been enhanced by 13$\%$ to roughly align peak heights.}
\end{figure}

In lower resolution studies, it is common practice to tune atomic multiplet models so that excitation features are centered on experimental peaks, and to account for much of the peak width using lifetime (Lorentzian) and experimental resolution (Gaussian) broadening. However, the high resolution measurements in Fig. 2(c) are not compatible with this approach. A broadened atomic multiplet calculation (Fig. 2(c), blue curve) cannot simultaneously match the sharp leading edge and higher energy structure on the energy loss side of the excitation.

The multiplet model is therefore used only to simulate the energetically sharp states at the leading edge of intensity, to better understand why the slope and leading edge energy change with temperature. Changing the crystal field parameter 10Dq by 1.9$\%$ as expected for thermal lattice expansion accounts for just approximately half of the onset energy shift (Fig. 2(c), green curve). Simultaneously weighting multiplet state occupancies with the 320K Boltzmann distribution reproduces the high temperature onset line shapes closely (Fig. 2(c), red curve and data points), as well as the drop in overall peak height \cite{SM}.

Thermal population of low energy modes softens the leading edge slope by allowing in-gap excitations. The first three excited states of the simulation are primarily associated with excitations of spin (20meV), spin orbit (38meV) and mixed (47meV) degrees of freedom respectively, as also discussed from perturbation theory in Ref. \cite{ElkeXLD}. These states have 22$\%$, 11$\%$ and 8$\%$ occupation respectively at 320K, and the ground state has only 45$\%$ occupation. When cooled to 90K, the system is almost fully in the ground state (93$\%$). Taken collectively, correlations with the resonance profile, excitation energies and temperature dependence strongly suggest that the atomic multiplet model is accurately describing electronic symmetries and energetics on the scattering site.

The multiplet scattering processes that lead to the thermal build-up of intensity on the energy gain side of a charge excited final state have striking similarity to anti-Stokes Raman emission. Anti-Stokes scattering processes annihilate thermally excited low energy modes and result in energy gain features in scattering spectra. A representative anti-Stokes process in multiplet calculations is drawn in Fig. 3(a), in which the system evolves from a spin excited initial state to the lowest energy state of the $^4T_1$ ground state manifold. In a many-body context, this is similar to the annihilation of a thermal magnon. In what we are calling pseudo-anti-Stokes (pAS) emission, the system evolves from a thermally excited initial state to the lowest energy state of a different multiplet state manifold (e.g. $^4T_2$), resulting in intensity on the energy gain side of a multiplet excitation. In a many-body context, the pAS scattering process drawn in Fig. 3(b) can be thought of as simultaneously annihilating a thermal magnon and creating an in-gap charge excitation (exciton).

The matrix elements and momentum dependence of pAS scattering are not identical to true anti-Stokes emission, but the close metaphor between the two provides an intuitive way to understand thermal changes throughout the simultated RIXS spectrum. To isolate electronic Boltzmann effects from the energy shift brought on by lattice relaxation, high and low temperature Boltzmann weighted simulations in Fig. 3(c) are performed with the same 10Dq crystal field parameter optimized for T=320K. Thermal broadening of all excitations is asymmetrical, with greater intensity on the energy gain side, as seen from the high and low temperature calculations in Fig. 3(c). The energy scale and fractional intensity of pAS intensity on the energy gain side of multiplet excitations are quite similar to true anti-Stokes emission for most of the CoO multiplet excitations.

\section{Interatomic shake-up and the excitation line shape}

The results presented in Fig. 2 show that atomic multiplet simulations accurately predict changes at the leading edge of the $^4T_2$ excitation. However, most of the asymmetric line shape of the $^4T_2$ excitation is unexplained, and must involve excited dynamics that have not yet been touched on. Because the line shape is not explained in a single atomic multiplet picture, it is necessary to consider many-body interactions with electronic degrees of freedom on the 12 nearest neighbor and 6 next neighbor cobalt atoms surrounding the scattering site. Interactions between the $^4T_2$ excitation and local many body degrees of freedom occur because the $^4T_2$ excitation involves a redistribution of electron density from $t_{2g}$ orbitals that point towards nearest neighbor Co atoms, into $e_g$ orbitals that overlap strongly with oxygen $p_\sigma$ orbitals. For Co atoms that neighbor the scattering site, this will lower the energy of d-orbitals that point towards the scattering site and enable magnetic exchange interactions through the vacated $t_{2g}$ state.

A fully interacting quantum model is not practical to implement with such a large cluster basis in a 3D crystal. We therefore neglect the entanglement of electronic states on neighboring Co atoms. In this approximation, all many-body eigenstates $\left|\Psi_\sigma\right>$ that incorporate multiple Co atoms are obtained by matrix multiplying over the single atom multiplet states:

\begin{eqnarray}
\left|\Psi_\sigma\right>=\left|\Psi_{0\sigma}\right>\prod_{i}\left|\Psi_{i\sigma}\right>
\end{eqnarray}

where $\left|\Psi_{0\sigma}\right>$ is an atomic multiplet eigenstate of the scattering site, and $\left|\Psi_{i\sigma}\right>$ represents an atomic multiplet eigenstate of a neighboring site, with i indexing the 18 nearest or next-nearest neighbor Co sites. The wavefunction of each Co site, $\left|\Psi_{i\sigma}\right>$, is determined by an independent calculation, with $\sigma$ indicating the magnetic exchange field orientation on the scattering site, and the magnetic orientation on other sites fixed by the antiferromagnetic order \cite{degOfFree}. The local Hamiltonian on neighboring sites is modified to account for different intersite energetics following the transfer of charge from $t_{2g}$ to $e_g$ symmetries in the $\sim$1eV $^4T_2$ excited states. Changes in magnetic exchange are described by adding an excitation-triggered exchange field $\tilde{B}$ oriented parallel to the magnetic orientation on the scattering site:

\begin{eqnarray}
\tilde{B}=1.5 \times n_{X} \sum_{i} J_\zeta,
\end{eqnarray}

Here $J_\zeta$ is the amplitude of the $^4T_2$ excitation-induced change in local magnetic interactions. The index $\zeta$ takes a value of 1 for the interaction with nearest neighbor Co sites and 2 for next nearest neighbors, consistent with the vectors labeled in Fig. 4(a). The prefactor of 1.5 represents the 3/2 spin moment. The operator $n_{X}$ has a value of 1 if a $\sim$1eV orbital excitation is present on the scattering site and 0 otherwise. Non-magnetic crystal field interactions are considered by adding a charge correlation Hamiltonian term $\tilde{V}$:

\begin{eqnarray}
\tilde{V}=n_{X} \sum_{i}n_{z_i^2}V_\zeta
\end{eqnarray}

where $n_{z_i^2}$ is an electron number operator acting on the $3d_{3z_i^2-r^2}$ symmetry orbital of the i indexed site. Spatial symmetry of the interaction is set by the $z_i$ coordinate axis, which is parallel to a vector connecting the indexed site to the scattering site. Linking the intersite interaction to $n_{z_i^2}$ is equivalent to treating the $^4T_2$ excitation as a point-source perturbation located along the axis connecting the Co scattering site to a neighboring Co site, which perturbs d-orbitals extending from the neighboring site.
\begin{figure}

\centering
\includegraphics[width = 8.7cm]{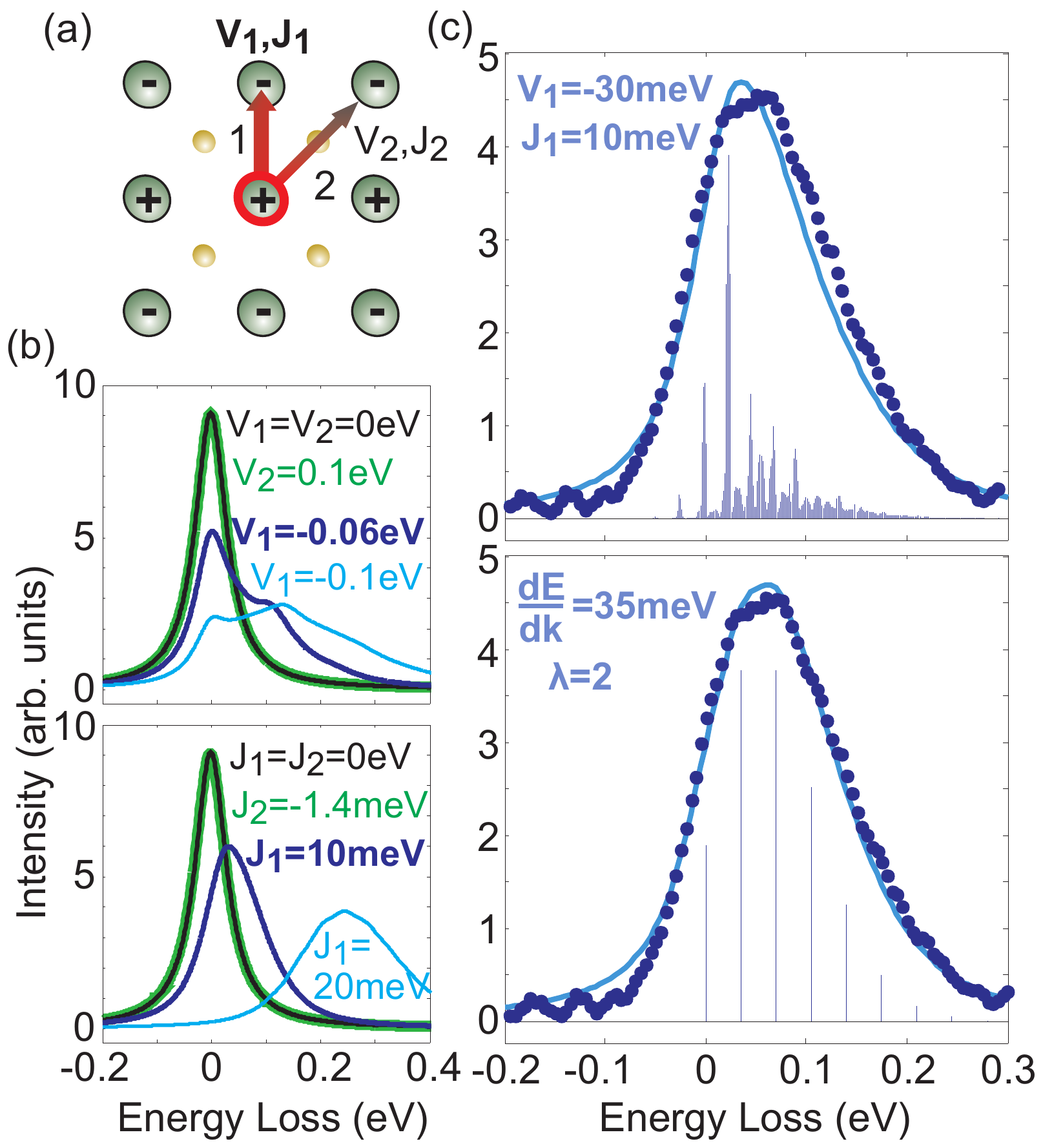}
\caption{{\bf{Intersite electronic shake-up interactions}}: (a) The scattering site is highlighted in red in a [001] plane of the CoO lattice, with relative antiferromagnetic spin orientations indicated by '+' and '-' signs. Parameters describing intersite interactions are labeled for (vector 1, [110] direction) nearest and (vector 2, [100] direction) next-nearest cobalt neighbors of the scattering site. (b) The effect of (top) V and (bottom) J shake-up parameters is simulated. Green curves representing nearest neighbor interactions (V$_2$=0.1eV and J$_2$=2.1meV) are nearly identical to (black) the trivial curves with no shake-up parameters applied. (c) Data points obtained by averaging the five T=90K curves in Fig. 2(c) are compared with (top) a shake-up simulation and (bottom) a Poisson distribution.}
\end{figure}

A shift in the $^4T_2$ excitation onset energy is avoided by setting the lowest energy atomic eigenstate on each neighbor of the scattering site to 0eV for both values of $n_{X}$ through an additional factor $\tilde{C}_X$. This is necessary because an average effect of the $\tilde{B}$ and $\tilde{V}$ interactions is implicitly included within the crystal field 10Dq parameter. The resulting Hamiltonian $\tilde{H}=\tilde{H}_{mult}+\tilde{V}+\tilde{B}+\tilde{C}_X$ builds on the multiplet Hamiltonian used for simulations in the previous section ($\tilde{H}_{mult}$), and acts on an effective Co$_{19}$O$_{44}$ cluster. The Hamiltonian is constructed such that all RIXS spectral features are unchanged except for the $\sim$1eV $^4T_2$ excitation. Each Co site has 12 low energy states, extending up to $\sim$140meV above the single-site ground state. Nonzero values of $J_\zeta$ and $V_\zeta$ allow the $^4T_2$ excitation to perturb low energy state symmetries on surrounding Co sites, giving a small probability that charge excitations will trigger low energy 0-140meV excitations on neighboring sites.

Calculations in Fig. 4 isolate the effect of $\tilde{B}$ and $\tilde{V}$ intersite interactions on the charge excitation line shape by treating the scattering site component of the $^4T_2$ excitation as a 0eV delta function. Several important trends are evident when the $V_\zeta$ and $J_\zeta$ terms are explored independently. First, $V_1$ has a far greater effect on the line shape than $V_2$, which may be understood because the $V_1$ perturbation directly disrupts the [100] ground state axis of rotation for orbital angular momentum. Second, the largest likely value of $J_2$, -1.4meV, is also insufficient to influence the excitation line shape. The $J_2$ parameter represents the suppression of oxygen mediated superexchange, which has a mean field interaction strength of 1.4meV seen from the antiferromagnetic exchange field (1.4meV=12.6meV/(6$\times$1.5)). Thus only the $V_1$ and $J_1$ have immediate importance in determining the excitation line shape.

The $J_1$ parameter is assigned larger amplitudes than $J_2$, because electron mobility and interactions between near neighbor Co atoms are thought to be much greater than oxygen mediated mobility between next neighbor Co sites \cite{CoO_LDA}. Either $J_1$ or $V_1$ can potentially explain the full width of the experimental line shape, however $J_1$ appears to give a better line shape correspondence with data and does not require strong interactions - just $J_1$$\sim$10meV is sufficient. The $V_1$ parameter is assigned a negative sign because it comes from a reduction in $t_{2g}$ charge density on the scattering site, but the qualitative effect of the $V_1$ perturbation is sign independent. The effect of the $J_1$ parameter does not depend at all on its sign, because the scattering site has an equal number of spin aligned and antialigned nearest neigbor Co sites. A comparison with T=90K RIXS data is presented in Fig. 4(c, top), using a balance of $V_1$ and $J_1$ intersite interaction parameters, and Boltzmann weighting the model to T=90K.

\begin{figure}
\centering
\includegraphics[width = 8.7cm]{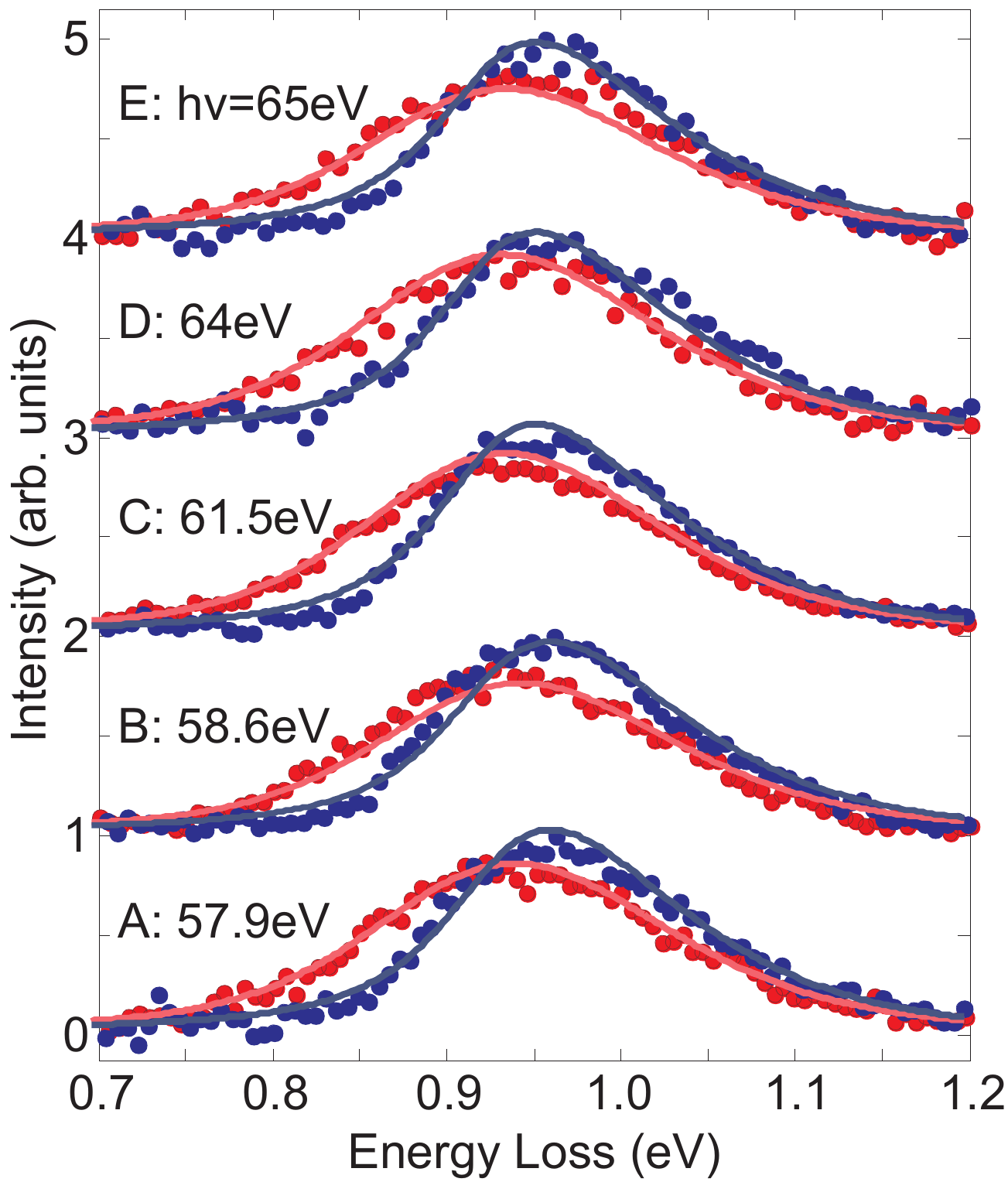}
\caption{{\bf{Merging shake-up and multiplet models}}: A simulation of the scattering site and its 12 nearest Co neighbors is thermalized to (dark blue curves) T=90K and (light red curves) T=320K to match experimental data. Data curves are reproduced from Fig. 2(c), and are measured at the incident energies labeled.}
\end{figure}

A meaningful analogy can be drawn between the probabilistic triggering of low energy excitations on atoms that neighbor a localized charge excitation and the Poisson distribution of high-repetition stochastic processes. When several excitations are likely to occur simultaneously, the simulated line shape tends to evolve into a Poisson-like (Gaussian-like) distribution as seen for the $J_1$=20meV simulation in Fig. 4(b, bottom). However, when there is unlikely to be more than one simultaneous excitation on neighbors of the RIXS scattering site, the spectrum can easily become irregular, with prominent bumps at the energies of specific low energy atomic multiplet excitations (Fig. 4(b, top)). When the calculation is fitted to our data in the Fig. 4(c, top), the average chance of triggering an excitation of any energy is only 6$\%$ per atom (Poisson $\lambda\sim2/3$), suggesting that both the stochastic physics of the Poisson distribution and the intrinsic $\sim$140meV distribution of low energy states are contributing factors to the $\sim$160meV experimental peak width. Fitting RIXS data with the Poisson distribution gives parameters of $\lambda$=2 (17$\%$ excitation chance/atom) and dE/dk=35meV, likely overstating the probability of low energy excitations occurring on the near neighbor Co atoms.

Full simulations that include the atomic multiplet states of the scattering site allow a comparison with the temperature and incident energy dependence of RIXS measurements (Fig. 5). Overall line shapes are matched well by the simulation, including the broader and smoother characteristic shape of T=320K data. Correspondence with temperature dependence at the leading edge of scattering intensity (0.8-0.9eV) remains quite good when the thermalized states of neighboring sites are considered, strongly supporting the analysis of Fig. 2(c). These results have great conceptual importance for time resolved pump-probe RIXS systems currently in development, because they show that RIXS can effectively reveal the thermal (or pumped/athermal) occupancy of low energy electronic degrees of freedom.

\section{Discussion}

Our model treats the multiplet state as a rapidly imposed classical perturbation on its surroundings, which is equivalent to regarding it as a quantum quench: a sudden change in the parameters of a system Hamiltonian, which creates new dynamics \cite{quenchDef}. Quantum quenches are theoretical abstractions that can only be approximated in experiments, and resolving the absolute accuracy of this picture will require future high resolution and momentum resolved studies. The idea that localized core electron excitations can cause a simple quantum quench that acts on neighboring atoms has been widely applied in models of ``shake-up" or ``indirect" resonant scattering processes \cite{Hasan_KRIXS,AmentRIXSReview,VDBultrashort,VDBmagneticLCO,DevereauxNew}. The present situation is unusual however in that the origin of the intersite perturbations is the atomic $^4T_2$ valence excitation, rather than a core hole resonance state. The Poisson-resembling line shape is thus intrinsic to well-resolved low energy orbital excitations in CoO, and not an artefact of the RIXS resonance process.

The model shows that quite weak electronic shake-up interactions can account for the observed line shape and temperature dependence, and provides approximate upper limits of $J_1$=10meV and $V_1$=100meV for the strength of those interactions. However, it is also possible that a measurable component of the shake-up intensity is due to coupling between the multiplet excitations and phonons. We have not specifically modeled the phonon shake-up component because high resolution RIXS studies have not yet resolved evidence for large coupling constants between phonons and charge neutral multiplet final states. Charge-bearing particles (e.g. electrons and holes) and transitions that change the principle quantum number of an electron (e.g. L- or K-edge resonance) cause all d-orbitals to expand or contract significantly in Hartree-Fock calculations, and are more natural candidates for strong phonon shake-up processes. In 1D materials without strong electronic localization, long-lived orbitons can have cleanly resolved momentum dispersion \cite{spinOrbitalSep}, which would be be obscured if the expected number of phonon scattering events were large (Poisson $\lambda\gtrsim1$). Coupling to electronic degrees of freedom can also be minimized by measuring RIXS from spectator atoms in a largely d$^8$ system, as high spin d$^8$ sites have fairly classical spin with only 3 low energy multiplet degrees of freedom that have weak excitation matrix elements with respect to the V- and J- type perturbtations considered in our model. The room temperature RIXS spectrum from Nd spectator sites in the largely $d^8$ nickelate system Nd$_{1.67}$Sr$_{0.33}$NiO$_{4+\delta}$ is compared with CoO RIXS data in Fig. 6. The electronically simple Nd f-orbital excitation is energetically sharp and does not reveal clear signs of energy loss features from phonon shake-up.

\begin{figure}
\centering
\includegraphics[width = 8cm]{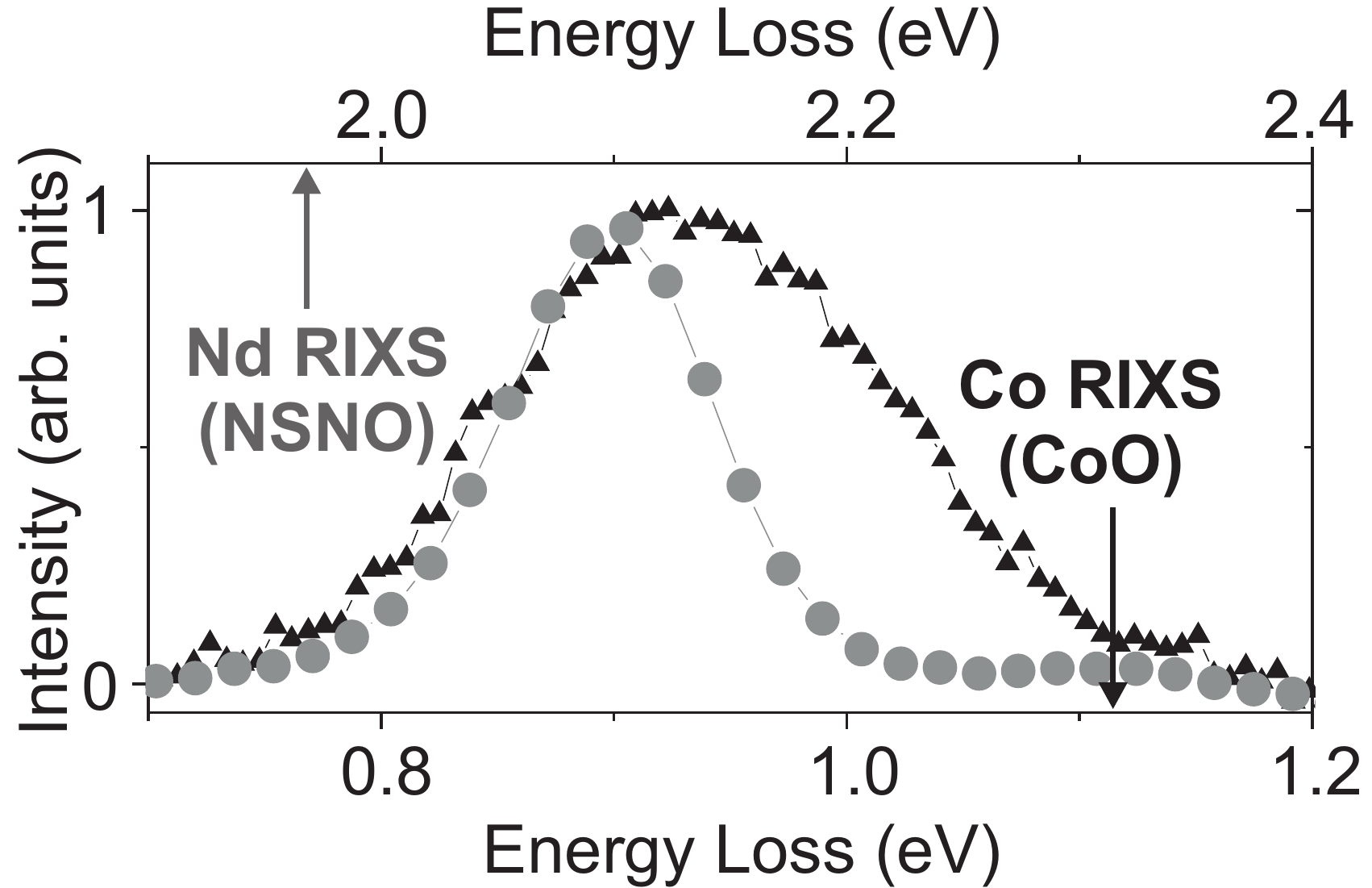}
\caption{{\bf{RIXS on a spectator atom}}: (black, diamonds) The CoO $^4T_2$ excitation measured at $h\nu$=61.5eV and T=320K is overlaid with (gray, circles) a room temperature $ff$ RIXS excitation spectrum from Nd spectator atoms in Nd$_{1.67}$Sr$_{0.33}$NiO$_{4+\delta}$ (NSNO). Energy loss for NSNO is displayed on the top axis, and is shifted by 1.2eV relative to the CoO incident energy axis.}
\end{figure}

Shake-up effects from intermediate core hole states cannot be responsible for the effect in our data, because RIXS curves excited via intermediate states with a $\Gamma_m$$\sim$0.5eV inverse lifetime, such as curve A, would lead to very different excitation line shapes than the intermediate states of curves D-E which have lifetimes of $\Gamma_m>2eV$ (from SCLC). Moreover, the intermediate state inverse lifetimes are much larger than the inverse dephasing times for low energy states ($\Delta$E$\lesssim$140meV), greatly limiting any intermediate state shake-up effect on the neighboring Co quantum states.

In summary, the model large spin Mott insulator CoO has been studied with high resolution RIXS, revealing a Poisson-like line shape that is unlike the Lorentzian (lifetime broadened) excitation contours in atomic multiplet models. Temperature dependence of an anti-Stokes-like tail on the energy gain side of the prominent $^4T_2$ feature is found to be consistent with the thermal population of low energy states, suggesting that high resolution RIXS can be used to `take the temperature' of a generic electron system. The energy loss contour of the excitation is explained by a poly-atomic model in which the excitation triggers a shake-up of low energy modes on neighboring Co atoms. Within the model, even quite weak interactions with a $^4T_2$ multiplet charge excitation are shown to transform its line shape by causing low energy excitations to appear stochastically on different atomic neighbors of the scattering site. Based on this, we anticipate that future high resolution studies may identify Poisson-like line shapes for highly localized charge excitations in other materials that have a large low energy density of states, such as many mid-transition metal oxides (e.g. cobaltates and manganites).

\textbf{Acknowledgements:}

The Advanced Light Source is supported by the Director, Office of Science, Office of Basic Energy Sciences, of the U.S. Department of Energy under Contract No. DE-AC02-05CH11231. Work at Brookhaven National Laboratory was supported by the U.S. Department of Energy under No. DEAC02-98CH10886.

\newpage


\section*{Supplementary material}

# Supplemental material: matrix elements and higher order modeling factors
(Dated: April 23, 2013)

This section addresses experimental calibration factors and higher order modeling considerations that may be of interest for future high resolution RIXS studies.

## CONTENTS

# EXPERIMENTAL CALIBRATION

Calibration of energy loss values across the CCD was performed separately for each measurement configuration (each incident energy) to ensure accuracy. Scattering intensity measured at each pixel was normalized by a roughly unitary factor representing the energy scale per pixel, which was non-uniform across the CCD, to ensure that intensity in the final spectra accurately represents the density of scattered photons per unit energy. Intensity was also normalized by the incident photon flux and the beamline/spectrometer efficiency vs. energy curves.

# SURFACE AND BULK LATTICE DISTORTION IN $CoO$

CoO undergoes a tetragonal lattice distortion that is expected to broaden lineshapes beneath 290K [1]. It appears empirically that this effect is small, because low temperature excitation onset lineshapes are quite sharp, and higher energy lineshapes are similar at high and low temperature. The tetragonal distortion has been approximated with a crystal field term of $D_S$=0.02eV at T=10K in recent literature [2], which is likely larger than occurs in the T=90K measurements here, and is greatly mitigated by the dominant L.L multiplet interaction terms. Including Ds=0.02eV causes splitting of just 11meV in leading edge states of the $^4T_2$ excitation, which is small relative to the experimental lineshape of 70meV.

A similar tetragonal lattice distortion will manifest intrinsically within a few nanometers of the crystal surface. The RIXS energy loss features from 4.2nm diameter CoO nanoparticles have recently been shown to be shifted to $\lesssim$10% lower energies relative to bulk CoO excitations [2]. Atoms near the surface termination of a large crystal exist in a far more bulk-like environment, however this trend suggests that surface relaxation could slightly broaden the energy gain side of M-edge RIXS excitations relative to intrinsic bulk line shapes. Strong surface effects are not expected, as the measured [010] face is non-polar and bulk-like properties (e.g. z-axis dispersion) are observed from photoemission on CoO with roughly 1/10th the measurement penetration depth of M-edge RIXS [3].

# THERMAL CHANGES IN INTENSITY MATRIX ELEMENTS

Because lattice expansion and electronic Boltzmann factors both affect the starting electronic state of the RIXS scattering process, they have a visible effect on the expected intensity of peak features which can also be used to identify if thermal changes in the electronic state are accurately captured by the model. Table I summarizes these matrix elements for the curves in Fig. 2 of the main text, and shows that the thermalized multiplet model reproduces the intensity change in all curves to within 3% statistical error.

|         | A    | B    | C    | D    | E    |
|---------|------|------|------|------|------|
| Data    | 0.88 | 0.87 | 0.86 | 0.90 | 0.82 |
| Model   | 0.86 | 0.84 | 0.89 | 0.90 | 0.82 |
| Lattice | 0.99 | 1.04 | 1.00 | 0.95 | 0.92 |

TABLE I. **Intensity matrix element change (90K→320K)**: The ratio of peak heights at T=320K and T=90K (I(320K)/I(90K)) for the data in Fig. 2(c) of the main text is compared with the thermalized multiplet model, and with a simulation of lattice distortion only.




\begin{thebibliography}
\bibitem{cubicNote} A tetragonal distortion beneath the N\'{e}el temperature is unlikely to be resolvable in our spectra \cite{SM}.
\bibitem{detailedVals} The inter-d-orbital and 3p-3d Slater-Condon parameters are set to 80$\%$ and 65$\%$ of atomic values (64$\%$ and 52$\%$ of Hartree-Fock). The d-orbital spin orbit coupling parameter is set to 0.06eV with no core hole and 0.07eV when a 3p core hole is present, and the 3p spin orbit coupling strength is set to 1.5eV.
\bibitem{degOfFree} A domain-dependent Z$_2$ degree of freedom in the spin orientation of nearest neighbor cobalt atoms is not relevant to the calculation results.
\end{thebibliography}
\end{document}